\documentclass[11pt]{article}

       \newcommand{\Dc}{ {\mathcal{D}} }

       \newcommand{\bfu}{{\mathbf u}}

       \newcommand{\bfx}{{\mathbf x}}

       \newcommand{\bfB}{{\mathbf B}}

       \newcommand{\bfJ}{{\mathbf J}}

  \newcommand{\nn}{\nonumber}
  \newcommand{\bsv}{\boldsymbol{v}}

\usepackage{graphicx}% Include figure files
\usepackage{epstopdf, epsfig}
\usepackage[colorlinks]{hyperref}
\usepackage{amsfonts}
\usepackage{amssymb}
\usepackage[T1]{fontenc}
\usepackage{mathtools}
\usepackage{amsfonts}
\usepackage{dcolumn}% Align table columns on decimal point
\usepackage{bm}% bold math
\usepackage[export]{adjustbox}
\usepackage{wrapfig}
\usepackage{lipsum}
\usepackage{authblk}
\allowdisplaybreaks
\usepackage[titletoc,toc,title]{appendix}	

\usepackage[style=ieee, citestyle=numeric-comp, backend=biber]{biblatex}

\usepackage{blindtext}
\usepackage[a4paper,
            bindingoffset=0.2in,
            left=1in,
            right=1in,
            top=1in,
            bottom=1in,
            footskip=.25in]{geometry}

\usepackage{hyperref}
\hypersetup{
    colorlinks=true,
    linkcolor=red,
    filecolor=magenta,      
    urlcolor=cyan,
    citecolor = cyan,
    }
 
\title{Axisymmetric hybrid Vlasov equilibria with applications to tokamak plasmas} 
\author[1,2]{D. A. Kaltsas \thanks{d.kaltsas@uoi.gr}}
\author[1]{A. Kuiroukidis} %\thanks{a.kuirouk@uoi.gr}}
\author[3]{P. J. Morrison} %\thanks{morrison@physics.utexas.edu}}
\author[1]{G. N. Throumoulopoulos} %\thanks{gthroum@uoi.gr}}
\affil[1]{Department of Physics, University of Ioannina, Ioannina GR 451 10, Greece}
\affil[2]{Department of Physics, Democritus University of Thrace, Kavala GR 654 04, Greece}
\affil[3]{Department of Physics and Institute for Fusion Studies,
The University of Texas at Austin, Austin, TX 78712, USA}

\date{}
 
\begin{document}
\maketitle
\begin{abstract}
We derive axisymmetric equilibrium equations in the context of the hybrid Vlasov model with kinetic ions and massless fluid electrons, assuming isothermal electrons and deformed Maxwellian distribution functions for the kinetic ions. The equilibrium system comprises a Grad-Shafranov partial differential equation and an integral equation. These equations can be utilized to calculate the equilibrium magnetic field and ion distribution function, respectively, for given particle density or given ion and electron toroidal current density profiles. The resulting solutions describe states characterized by toroidal plasma rotation and toroidal electric current density. Additionally, due to the presence of fluid electrons, these equilibria also exhibit a poloidal current density component. This is in contrast to the fully kinetic Vlasov model, where axisymmetric Jeans equilibria can only accommodate toroidal currents and flows, given the absence of a third integral of the microscopic motion.
\end{abstract}

%\twocolumn

\section{Introduction}
\label{Sec_I}
Hybrid Vlasov models play an important role in examining the complex behavior of multiscale plasmas that feature both a fluid bulk and energetic particle populations not amenable to fluid descriptions.  One specific branch of hybrid models that has received significant attention, primarily for studying phenomena in ion inertial scales such as turbulence and collisionless reconnection, focuses on electron-ion plasmas where electrons are treated as a fluid while ions are treated kinetically (e.g. \cite{Valentini2007, Servidio2012,Servidio2015, Cerri2016, Gonzalez2020,Tenerani2023, Winske2003, Le2016}). In our recent work (\cite{Kaltsas2023}), we employed such a hybrid model, featuring massless isothermal electrons and kinetic ions, to investigate one-dimensional Alfv\'en-BGK (Bernstein-Greene-Kruskal) modes as stationary solutions to the model equations. We demonstrated that the one-dimensional equilibrium equations constitute a Hamiltonian system for a pseudoparticle, which can exhibit integrable or chaotic orbits, depending on the form of the distribution function. A natural extension of this work would be the construction of 2D-equilibria which can be used as reference states for studying reconnection, instabilities and wave propagation, or even macroscopic equilibria of fusion plasmas. 

Plasmas in fusion devices like the tokamak, are enriched with significant populations of energetic particles. It is thus expected that the distribution of those particles in the physical and the velocity space might affect macroscopic equilibrium and stability properties. For this reason hybrid models have also found applications in the description of multiscale dynamics of tokamak plasmas (e.g. \cite{Park1999, Todo2016, Todo2021}).  However, despite the utility of hybrid and kinetic descriptions for investigating dynamical processes, there has been limited progress in constructing self-consistent equilibria within the framework of these models. One important limitation arises from the absence of a third particle constant of motion in the full-orbit Vlasov description. Such an invariant would be crucial for constructing equilibria with characteristics relevant to tokamaks. Efforts to build such equilibria using a fully kinetic Vlasov description for both ions and electrons have been undertaken in \cite{Tasso2014, Kuiroukidis2015}. Nevertheless, due to the presence of only one momentum integral of motion for each particle species, specifically the particle toroidal angular momentum, these equilibria exhibit only toroidal current density and plasma rotation. In contrast, the magnetohydrodynamic (MHD) fluid description of toroidal plasma equilibrium can accommodate both toroidal and poloidal currents. Hence, although more fundamental, the kinetic approach appears to have limitations in describing certain classes of equilibria compared to MHD. 

To combine the advantages of both descriptions, we turn to the hybrid model mentioned earlier. Even though it lacks a poloidal particle momentum invariant, it can describe equilibria featuring both toroidal and poloidal current densities, due to the fluid treatment of electrons, which carry the poloidal current component. Considering kinetic ions and fluid electrons might be relevant to tokamak scenarions with significantly higher ion than electron temperatures known as hot-ion-modes achieved by injecting highly energetic neutral beams (e.g. \cite{McNamara2023}).  It is important to note though that a limitation of the present model for a realistic description of fusion plasmas, is that it treats the entire ion population using the Vlasov equation. This approach rules out the possibility of macroscopic poloidal ion flows and the existence of multiple ion species; thus further model improvements are required. The present model description serves as an initial step toward the development of improved models that will incorporate multi-fluid-kinetic descriptions, as exemplified in \cite{Kaltsas2021}.

The rest of the paper is structured as follows: in Section \ref{Sec_II} we present the hybrid equilibrium model and in Section \ref{Sec_III} the axisymmetric equilibrium formulation is developed. In Section \ref{Sec_IV} we numerically construct particular tokamak-pertinent equilibria presenting various equilibrium characteristics and we conclude by summarising the results in Section \ref{Sec_V}.

\section{The hybrid model}
\label{Sec_II}
The initial hybrid-Vlasov equilibrium system employed in \cite{Kaltsas2023}, consists of a Vlasov equation for kinetic ions, a generalized Ohm's law derived from the electron momentum equation assuming massless electrons, the Maxwell equations, and an equation of state for the fluid electrons:
\begin{align}
    \bsv \cdot\nabla f +\frac{e}{m}\left(\textbf{E} + \bsv \times \textbf{B}\right)\cdot\nabla_{\bsv} f =0 \,,\label{Vlasov}\\
    \textbf{E} = -\frac{n}{n_e}\textbf{u}\times \bfB + \frac{\bfJ \times  \bfB }{en_e} -\frac{\nabla P_e}{en_e} \,,\label{Ohm}\\
    \textbf{E} = -\nabla\Phi\,, \quad \nabla \times \bfB = \mu_0 \bfJ\,, \\
    \quad \nabla\cdot\bfB=0\,, \quad \nabla\cdot\textbf{E} = \frac{e}{\epsilon_0}(n-n_e)\,, \\
    P_e = n_e k_B T_e\,, \label{EOS}
\end{align}
where
\begin{eqnarray}
    n(\bfx,t)= \int d^3v\, f(\bfx,\bsv,t)\,,\label{n_integral}\\
    \bfu(\bfx,t) = n^{-1}\int d^3v\, \bsv f(\bfx,\bsv,t)\,. \label{u_integral}
\end{eqnarray}
Here $f(\textbf{x},\bsv,t)$ is the ion distribution function. Note that the ion-kinetic contribution to the current density is given by
\begin{eqnarray}
    \textbf{J}_k = \int d^3v\, \bsv f \,, \label{Jk}
\end{eqnarray}
and thus the first term in the right hand side of \eqref{Ohm} can be expressed as $-\textbf{J}_k\times \textbf{B}/(en_e)$.

In addition to \eqref{Vlasov}--\eqref{EOS}, an energy equation is needed to determine  $T_e$. Alternatively, it can be assumed that the electrons are isothermal, i.e. $T_e$ is constant throughout the entire plasma volume, or it can vary with the magnetic flux function $\psi$ if we consider isothermal magnetic surfaces, i.e., $T_e = T_e(\psi)$. An alternative to \eqref{EOS} would be an isentropic closure of the form $P_e = c n_e^\gamma\,,$ or even anisotropic electron pressure under appropriate conditions for the different components of the electron pressure tensor. Here we consider isothermal electrons $T_e=T_{e0}=const.$

Let us now write the system \eqref{Vlasov}--\eqref{EOS} in nondimensional form upon introducing the following dimensionless quantities
\begin{eqnarray}
&&\tilde{x}=\frac{x}{R_0}\,, \quad \tilde{\bsv}=\frac{\bsv}{d_i v_{A}} \,, \nn \\
&& \tilde{n}_e=\frac{n_e}{n_0}\,, \quad
\tilde{f}=d_i^3 v_{A}^3 f/n_0\,, \nn \\
&&\tilde{\textbf{E}}=\frac{\textbf{E}}{d_i v_AB_0}\,, \quad \tilde{\textbf{B}}=\frac{\textbf{B}}{ B_0}\,,\nn\\
&&\tilde{\textbf{J}}_k=\frac{\textbf{J}_k}{en_0 d_i v_{A}}\,, \quad \tilde{P}_e=\frac{P_e}{m n_0 d_i^2 v_A^2}\,, \label{normalization}
\end{eqnarray}
where $R_0$ and $B_0$ are the characteristic length and magnetic field modulus, respectively. Additionally,
\begin{eqnarray}
v_A = \frac{B_0}{\sqrt{\mu_0 m n_0}}\,, \quad
\Omega = \frac{eB_0}{m}\,,
\end{eqnarray}
are the Alfv\'en speed and the ion cyclotron frequency, respectively and
\begin{eqnarray}
    d_i = \frac{\ell_i}{R_0}\,, \quad \ell_i = \sqrt{\frac{m}{\mu_0 n_0 e^2}}\,, 
\end{eqnarray}
is the nondimensional ion skin depth which is typically of the order $10^{-2}$ in fusion devices. Notice that apart from nondimensionalizing various physical quantities, we've also scaled the nondimensional velocity by a factor of $d_i^{-1}$. The rationale behind this scaling will be clarified in a subsequent explanation. What is important to stress here, is that with careful implementation of this scaling process, there are no inconsistencies in the nondimensionalization of the equations and the recovery of physical units in the final results. In view of \eqref{normalization} the hybrid equilibrium system then can be written in the following nondimensional form:
\begin{eqnarray}
    \bsv\cdot\nabla f + d_i^{-2}\left(\textbf{E} + \bsv\times \textbf{B} \right)\cdot\nabla_v f=0 \,, \label{Vlasov_nondim} \\
    -\nabla \Phi = \frac{1}{n_e}\left[\left(\nabla \times \textbf{B} - \textbf{J}_k\right) - d_i^2\nabla P_e\right]\,, \label{Ohm_nondim}\\
    \textbf{E} = - \nabla \Phi\,, \quad \nabla \times \textbf{B} = \textbf{J}\,,\\
    \nabla\cdot\textbf{B} = 0\,, \quad d_i^2\beta_A^2 \nabla\cdot E =(n-n_e)\,, \\
        P_e = \kappa n_e\,, \label{EOS_norm}
\end{eqnarray}
where 
\begin{eqnarray}
\kappa := \frac{k_B T_{e0}}{d_i^2mv_A^2}\,. \label{kappa}
\end{eqnarray} 
and $\beta_A^2 = v_A^2/c^2$. Taking the limit $\beta_A^2 \rightarrow 0$ we obtain the quasineutrality condition $n_e=n$, which will be applied in the subsequent analysis. 

In Section \ref{Sec_III}, we will investigate two equilibrium scenarios: one with cold electrons ($\kappa =0$) and the other with thermal electrons ($\kappa \sim 1$). We opted for the scaled nondimensional particle velocity $\tilde{v} = v/(d_i v_A)$ with the goal of attaining tokamak-relevant temperatures assuming $\kappa \sim 1$. It can be verified through \eqref{kappa} and using an Alfv\'en speed calculated from tokamak-relevant values for the density and the magnetic field, that  $\kappa \sim 1$ corresponds to $T_{e0}\sim 10^8\,K$. 

As a closing note for this section, it is worth highlighting that taking into account \eqref{EOS_norm} and the quasineutrality condition $n_e=n$, Ohm's law \eqref{Ohm_nondim} can be expressed as follows:
\begin{eqnarray}
    -\nabla \Phi = \frac{1}{n} \left(\nabla\times \textbf{B} - \textbf{J}_k\right)\times \textbf{B} - \nabla ln \left(n^{d_i^2\kappa}\right)\,. \label{Ohm_isothermal}
\end{eqnarray}

\section{Axisymmetric equilibrium formulation}
\label{Sec_III}
We consider a plasma configuration with axial symmetry with respect to a fixed axis, where all quantities depend on the coordinates $r,z$ of a cylindrical coordinate system $(r,\phi,z)$. Note that $z$ coincides with the axis of symmetry.  In this case the divergence-free magnetic field can be written in terms of two scalar functions $I$ and $\psi$ as follows:
\begin{align} 
    \bfB = I(r,z) \nabla \phi + \nabla\psi(r,z)\times\nabla\phi\,, \label{B_axisym}
\end{align}
while the corresponding current density is 
\begin{eqnarray}
        \bfJ = \nabla \times \bfB= -\Delta^*\psi \nabla \phi +\nabla I \times \nabla \phi\,, \label{J_axisym}
\end{eqnarray}
where  $\Delta^*$ is the Shafranov operator given by
\begin{equation}
    \Delta^* = r \frac{\partial}{\partial r} \left(\frac{1}{r} \frac{\partial}{\partial r}\right)+\frac{\partial^2}{\partial z^2}\,.
\end{equation}
Next we will consider the three components of \eqref{Ohm_isothermal} along the magnetic field, along the $\hat{\phi}$ direction and along the $\nabla \psi$ direction. From the $\textbf{B}$ projection we readily obtain
\begin{eqnarray}
    \textbf{B}\cdot \nabla \left[\Phi - ln \left(n^{d_i^2\kappa}\right) \right]=0\,,
\end{eqnarray}
thus 
\begin{eqnarray}
    \Phi - ln \left(n^{d_i^2\kappa}\right) \eqqcolon G(\psi)\,,
\end{eqnarray}
where $G(\psi)$ is an arbitrary function.  From this equation we can solve for $n$ to find
\begin{eqnarray}
    n = exp\left[\frac{\Phi-G(\psi)}{d_i^2\kappa}\right]\,. \label{n_Phi_G}
\end{eqnarray}
In the case $G(\psi)= const.$ we recover the Boltzmann distribution. Next, to take the $\nabla \phi$ and $\nabla \psi$ projections we  need first to determine the direction of $\textbf{J}_k$. 

According to Jeans' theorem \cite{Jeans1915,Lynden-Bell1962a}, distribution functions of the form $f=f(C_1,C_2,...)$, where $C_i$ are particle constants of motion, are solutions to the Vlasov equation \eqref{Vlasov_nondim}. In the absence of collisions, the particle energy $H$ is itself a first integral of motion. In  nondimensional form $H$ reads:
\begin{eqnarray}
    \tilde{H} = \frac{v^2}{2} + d_i^{-2}\Phi\,,
\end{eqnarray}
where $\tilde{H} = H/ (d_i^2m v_A^2)$. Additionally, in the presence of axial symmetry a second constant of motion is the particle toroidal angular momentum
\begin{eqnarray}
    \tilde{p}_\phi = r v_\phi + rA_\phi = rv_\phi + d_i^{-2}\psi\,,
\end{eqnarray}
where $\tilde{p}_\phi=p_\phi/(mR_0 d_i v_A)$. It remains an open question whether and under what conditions, additional, approximate constants of motion exist within the framework of full-orbit Vlasov description (see \cite{Efthymiopoulos2007} and references therein for a discussion on the existence of an approximate third integral of motion in axisymmetric potentials). In certain scenarios, it may be pertinent to consider adiabatic constants, such as the magnetic moment $\mu$ as explored in \cite{Aibara2021}. It is worth noting that in the context of the hybrid model and the present analysis, some assumptions made in \cite{Aibara2021}, such as $p_\phi \approx \psi$, can be justified due to the presence of the significant $d_i^{-2}$ factor, especially in systems like the magnetosphere. However, in this paper, which focuses on laboratory plasmas, we will not adopt this assumption. Instead, we will follow the approach outlined in \cite{Kaltsas2023} and \cite{Allanson2016}, considering a distribution function in the form of:
\begin{eqnarray}
    f = exp(-H) g(p_\phi) = exp\left[ - \frac{v_r^2+v_z^2+v_\phi^2}{2} - d_i^{-2}\Phi(r,z)\right]g(p_\phi)\,,
\end{eqnarray}
or 
$$ f= exp\left[ -\frac{v_r^2 +v_z^2}{2} - \frac{(p_\phi-\psi)^2}{2r^2} - d_i^{-2}\Phi(r,z)\right]g(p_\phi)\,. $$
Note that the tildes have been omitted in $H$ and $p_\phi$ for  convenience.
For such a distribution function the kinetic current density \eqref{Jk} will have only a $\phi$ component. This is because $v_r f$ and $v_z f$ are odd functions with respect to $v_r$ and $v_z$ respectively, while the integration over these variables goes from $-\infty$ to $+\infty$. Therefore
$$\textbf{J}_k = rJ_{k\phi}\nabla \phi\,,$$
and as a result
\begin{eqnarray}
\textbf{J}_{k}\times \textbf{B}= \frac{J_{k\phi}}{r}\nabla \psi\,. \label{Jkphi}
\end{eqnarray}
Substituting \eqref{B_axisym}, \eqref{J_axisym}, \eqref{n_Phi_G} and \eqref{Jkphi} into \eqref{Ohm_isothermal} we obtain
\begin{eqnarray}
    -n \nabla \Phi = -\frac{\Delta^* \psi}{r^2}\nabla \psi - \frac{I}{r^2}\nabla I +[I,\psi]\nabla \phi - \frac{J_{k\phi}}{r}\nabla\psi - n \nabla \Phi + n G'(\psi)\nabla\psi\,, \label{Ohm_axisym}
\end{eqnarray}
where $[a,b] \coloneqq (\nabla a \times \nabla b)\cdot \nabla \phi$. It is now trivial to see that from the $\nabla \phi$ projection of \eqref{Ohm_axisym}, we obtain
\begin{eqnarray}
    [I,\psi] =0 \,, \quad \text{i.e.\,} \quad I=I(\psi)\,.
\end{eqnarray}
Finally, the $\nabla\psi$ projection yields
\begin{eqnarray}
   \Delta^*\psi + II'(\psi) + r^2\mathcal{Z}(r,\psi)=0\,, \label{GS_1}
\end{eqnarray}
where $$\mathcal{Z}(r,\psi):= \frac{1}{r} \int d^3v\, v_\phi f -  G'(\psi) \int d^3v\, f\,.$$
Equation \eqref{GS_1} is a Grad-Shafranov (GS) equation determining the magnetic field through the flux function $\psi$ in axisymmetric hybrid Vlasov equilibria. Let us now work out the velocity space integrals in \eqref{GS_1}. The particle density is 
\begin{eqnarray}
    n = \int d^3v \, f = \frac{e^{-\Phi/d_i^2}}{r}\int_{-\infty}^{+\infty} dv_r \int_{-\infty}^{+\infty} dv_z \int_{-\infty}^{+\infty} dp_\phi\, e^{-\frac{v_r^2}{2}- \frac{v_z^2}{2} - \frac{(p_\phi-\psi/d_i^2)^2}{2r^2}}g(p_\phi)\nonumber\\=\frac{2\pi e^{-\Phi/d_i^2}}{r}\int_{-\infty}^{\infty}dp_\phi\, e^{-\frac{(p_\phi-\psi/d_i^2)^2}{2r^2}}g(p_\phi)\,.
\end{eqnarray}
We have shown that  $\Phi = ln (n^{d_i^2\kappa}) +G(\psi) $, therefore 
\begin{eqnarray}
    n = \left[\frac{2\pi e^{-G(\psi)/d_i^2}}{r} \int_{-\infty}^{+\infty}dp_\phi\, e^{-\frac{(p_\phi-\psi/d_i^2)^2}{2r^2}}g(p_\phi)\right]^{\frac{1}{\kappa+1}}\,. \label{n_final}
\end{eqnarray}
Similarly, for the toroidal component of the kinetic current density we find
\begin{eqnarray}
    J_{k\phi} = 2\pi e^{-G(\psi)/d_i^2}\left[\frac{2\pi e^{-G(\psi)/d_i^2}}{r} \int_{-\infty}^{+\infty}dp_\phi\, e^{-\frac{(p_\phi-\psi/d_i^2)^2}{2r^2}}g(p_\phi)\right]^{-\frac{\kappa}{\kappa+1}}\times \nn \\
    \times\int_{-\infty}^{+\infty} dp_\phi \, \frac{(p_\phi-\psi/d_i^2)}{r^2}e^{-\frac{(p_\phi-\psi/d_i^2)^2}{2r^2}}g(p_\phi)\,.
\end{eqnarray}
Therefore, the current density depends on two arbitrary functions, i.e. $G(\psi)$ and $g(p_\phi)$. The latter function that determines the ion distribution function, can either be specified a-priori, together with $G(\psi)$ and then the GS equation \eqref{GS_1} can be solved to determine $\psi$, or can be identified by fixing $n$ or $J_{k\phi}$ and $G(\psi)$. Following the formalism of \cite{Kaltsas2023} we can show that the function $\mathcal{Z}(r,\psi)$ can be derived by a ``pseudopotential'' function $V(r,\psi)$ which takes the form
\begin{eqnarray}
    V(\psi, r) = d_i^2(\kappa +1 )\left[\frac{2\pi e^{-G(\psi)/d_i^2}}{r} \int_{-\infty}^{+\infty} dp_\phi\, e^{-\frac{(p_\phi-\psi/d_i^2)^2}{2r^2}}g(p_\phi)\right]^{\frac{1}{\kappa+1}}= d_i^2(\kappa+1 ) n\,.\label{integral_V}
\end{eqnarray}
We can easily verify that 
$\mathcal{Z} = \frac{\partial V}{\partial \psi}\,,$
thus, the Grad-Shafranov equation can be written in the familiar form
\begin{eqnarray}
    \Delta^*\psi + II'(\psi) + r^2\frac{\partial V}{\partial \psi}=0\,. \label{GS_2}
\end{eqnarray}

Note that equation \eqref{GS_2} is reminiscent of the MHD GS equation with toroidal flow \cite{Maschke1980}, where an effective pressure function associated with the thermodynamic pressure and the plasma flow, appears instead of $V(r,\psi)$.  

To solve Eq.~\eqref{GS_2} we can specify $V(\psi,r)$ to be a known mathematical function or it can potentially be inferred by experimental data for the particle density $n$ or the toroidal current density profile. The feasibility of the latter approach will be explored in future work. Note that the particle density and the total toroidal current density can be expressed in terms of $V$ as follows 
\begin{eqnarray}
    n = \frac{V}{d_i^2(\kappa+1)}\,, \quad
    J_{\phi} = \frac{II'}{r} + r \frac{\partial V}{\partial \psi}.
\end{eqnarray}
Also note that that the electron contribution to $J_\phi$ is given by
\begin{eqnarray}
    J_{e\phi} = \frac{II'}{r} - r n G'(\psi)\,.
\end{eqnarray}
Knowing $V$ enables the solution of the partial differential equation \eqref{GS_2} to determine $\psi$ and of the integral equation \eqref{integral_V} to determine $g(p_\phi)$.

 In the Appendix \ref{appendix_A} we demonstrate that when the product $V^{\kappa+1} e^{G(\psi)/d_i^2}$ can be expressed as a power series expansion of $\psi$, it becomes possible to determine the function $g(p_\phi)$ in terms of Hermite polynomials. The function $g(p_\phi)$ is essentially determined upon calculating the coefficients  $c_m$ appearing in the expansion \eqref{g_Hermite} of the Appendix \ref{appendix_A}. This is done by solving Eq.~\eqref{V=sum-Hermite_3} for $c_m$ after expressing the left hand side (lhs) of \eqref{V=sum-Hermite_3} as a power series expansion in $\psi$. In this work we consider the special case
\begin{eqnarray}
    \left[\frac{V}{d_i^2(\kappa+1)}\right]^{\kappa+1} e^{G(\psi)/d_i^2}=V_0(r) +V_1(r)\psi + V_2(r)\psi^2\,, \label{ansatz_V}
\end{eqnarray}
and deal with two classes of equilibria corresponding to cold electrons with $\kappa=0$ and thermal electrons with $\kappa \sim 1$.   We consider here the quadratic  ansatz  \eqref{ansatz_V}  due to its simplicity in determining the constants $c_m$ in the expansion \eqref{V=sum-Hermite_3}. While higher-order terms in $\psi$ could be included, such an extension would complicate the analysis, exceeding the scope of the present study, which aims to showcase the feasibility of solving the inverse problem via the Hermite polynomial method. Another motivation for this choice is that adopting a similar ansatz for the free function $I^2(\psi)$ and assuming $G(\psi)=0$, results in a linear Grad-Shafranov equation in the cold electron limit, which is the simpler case with non-monotonic current density profile and can be solved analytically in terms of truncated power series as done in \cite{Kaltsas2019c}. 

By equations \eqref{ansatz_V} and \eqref{V=sum-Hermite_3} we see that 
\begin{eqnarray}
    V_0 + V_1 \psi + V_2 \psi^2 = c_0 2\pi^{3/2} + c_1 \frac{(2 \pi)^{3/2}}{d_i^2} \psi + c_2 \frac{4\pi^{3/2}}{d_i^4}\psi^2 + c_2 4 \pi^{3/2} (r^2-1)\,, 
\end{eqnarray}
and therefore the coefficients $c_0$, $c_1$ and $c_2$ are 
\begin{eqnarray}
    c_0 = \frac{V_0(r)}{2\pi^{3/2}}- 2 c_2(r^2-1)\,, \quad
    c_1 = \frac{d_i^2V_1}{2\pi^{3/2}}\,, \quad
    c_2 = \frac{d_i^4 V_2}{4\pi^{3/2}} \,.
\end{eqnarray}
In order for $c_0,c_1,c_2$ to be constants we should select $V_1=const.$ $V_2=const.$ and 
$$V_0(r) = C_0 +d_i^4 V_2 (r^2-1)\,,$$
where $C_0$ is a constant. Therefore, the ion distribution function in both the cold and thermal electron limits
reads as follows:
\begin{eqnarray}
    f(H,p_\phi) = \left[ c_0+\sqrt{2}c_1p_\phi+c_2(2p_\phi^2-2)\right]e^{-H}\,. \label{DF_special}
\end{eqnarray}
To ensure the positivity of the distribution function \eqref{DF_special} it suffices to require $c_0+\sqrt{2}c_1p_\phi+c_2(2p_\phi^2-2)>0$, $\forall p_\phi$, which holds true for
$$c_0 > \frac{c_1^2 - 8 c_2^2}{4c_2}\,, \quad c_2>0\,.$$

\section{Tokamak equilibria}
\label{Sec_IV}

To fully define  the plasma equilibria we further need to specify the free functions $I(\psi)$ and $G(\psi)$. In this work we adopt 
\begin{eqnarray}
    I^2(\psi) &=& (I_0+I_1\psi +I_2 \psi^2)e^{-\eta(\psi -\psi_{a})^2}\,, \label{ansatz_I} \\
    G(\psi) &=& \alpha(\psi-\psi_a)^2\,. \label{ansatz_G}
\end{eqnarray}
Here, $I_0$, $I_1$, $I_2$, $\eta$, and $\alpha$ are constants, and $\psi_a$ represents the value of the flux function $\psi$ at the magnetic axis, corresponding to an elliptic O-point of $\psi$ where the magnetic field is purely toroidal.  This particular choice for $G(\psi)$ is  based on the expectation that the number density should decrease towards the plasma boundary, (see Eq.~\eqref{n_final}). This choice also plays a crucial role in generating flow and toroidal current shear in the edge region of the plasma.

The specific form of the ansatz \eqref{ansatz_I} for $I^2(\psi)$ is influenced by the structure that $V$ assumes after adopting the forms given in \eqref{ansatz_V} and \eqref{ansatz_G}. Furthermore, this ansatz offers substantial flexibility in shaping various equilibrium profiles although the examples of equilibria provided below were obtained for specific values of the free parameters, rather than through a detailed exploration of the parametric space.

We address the fixed-boundary equilibrium problem with a tokamak-relevant, D-shaped computational domain denoted as $\mathcal{D}$. In this context, we solve the Grad-Shafranov equation \eqref{GS_2} while specifying $V$ as
\begin{eqnarray}
    V = d_i^2 e^{-G(\psi)/d_i^2}\left[V_0(r) + V_1 \psi +V_2 \psi^2 \right]\,, \label{ansatz_V_cold}\\
    V = 2 d_i^2 \left\{e^{-G(\psi)/d_i^2} \left[V_0(r) + V_1 \psi +V_2 \psi^2  \right]\right\}^{1/2}\,, \label{ansatz_V_hot}
\end{eqnarray}
for the $\kappa=0$ and $\kappa=1$ cases, respectively. It should be noted that it is possible to calculate thermal electron equilibria with different values of $\kappa$ rather than $\kappa=1$. This would correspond to assuming a different electron temperature and also the expression in Eq.~\eqref{ansatz_V_hot} would vary accordingly. However, $\kappa=1$ corresponds to a tokamak-relevant temperature, making it both a convenient and experimentally relevant choice. Another interesting possibility is allowing $\kappa$ to depend on the flux function $\psi$, resulting in equilibria with isothermal magnetic surfaces. This will be explored in a future study.  The boundary condition is of Dirichlet type given by $\psi|_{\partial \mathcal{D}}=0$ where $\partial \Dc$ is the boundary of a computational domain $\mathcal{D}$. This corresponds to a closed flux surface embedded into the plasma rather than the actual plasma boundary as it remains unclear how to determine the appropriate conditions that will produce the desired profile behavior in the outermost plasma region within the kinetic framework (see also the relevant discussion in \cite{Kuiroukidis2015}). For cold electrons the Grad-Shafranov equation \eqref{GS_2} takes the familiar form
\begin{eqnarray}
    \Delta^*\psi + II'(\psi)+ d_i^2 e^{-G/d_i^2}\left[\left(V_1 +2 V_2 \psi \right)-\frac{G'(\psi)}{d_i^2}\left(C_0-d_i^4V_2+V_1\psi + V_2\psi^2\right)\right]r^2\nn\\
    - d_i^2V_2 e^{-G/d_i^2}G'(\psi) r^4=0\,.  
\end{eqnarray}
Note that a Grad-Shafranov equation of similar structure describes axisymmetric equilibria with incompressible flows of arbitrary direction, as shown in \cite{Tasso1998}. In the MHD context the $r^4$ term is associated with the non-parallel to the magnetic field component of the flow.

\begin{figure}[h!]
    \centering
    \includegraphics[scale=0.65]{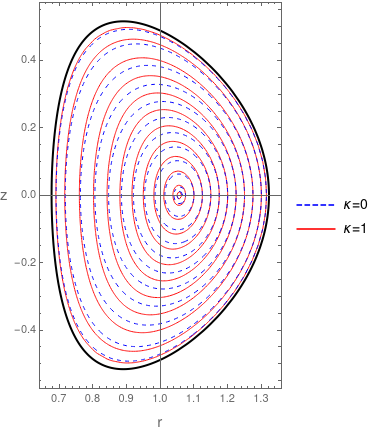}
    \caption{Magnetic surfaces of an equilibrium with cold electrons (blue dashed lines) and an equilibrium with thermal electrons (red solid lines).}
    \label{fig_contours}
\end{figure}

\begin{figure}[h!]
    \centering
    \includegraphics[scale=0.45]{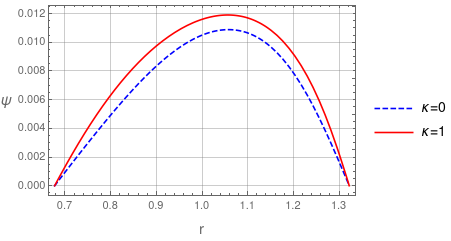}
    \includegraphics[scale=0.45]{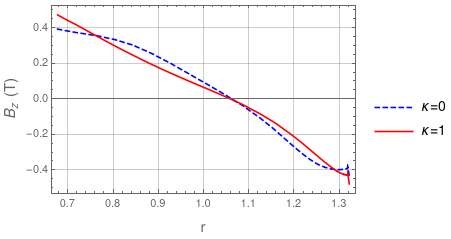}
    \caption{Variation of the flux functions $\psi$ (left) and the $z$ component of the magnetic field (right) along the $r$ axis on the equatorial plane $z=0$. The dashed blue lines correspond to the cold electron equilibrium and the red solid lines correspond to thermal electrons.}
    \label{fig_u_Bz}
\end{figure}

\begin{figure}[h!]
    \centering
    \includegraphics[scale=0.45]{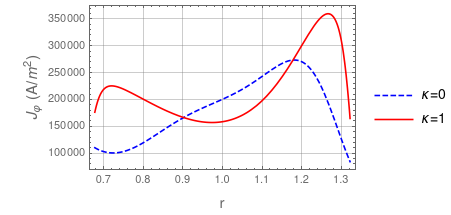}
    \includegraphics[scale=0.45]{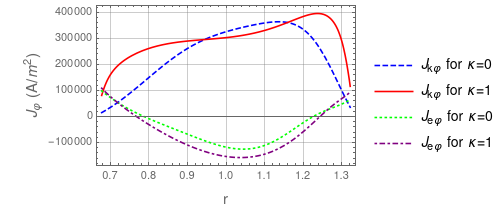}
    \caption{The toroidal current density profiles for the two equilibrium classes. The total current density profile is displayed in the left panel, while in the right panel the electron and the ion kinetic contributions are drawn separately.}
    \label{fig_J_profiles}
\end{figure}

\begin{figure}[h!]
    \centering
    \includegraphics[scale=0.45]{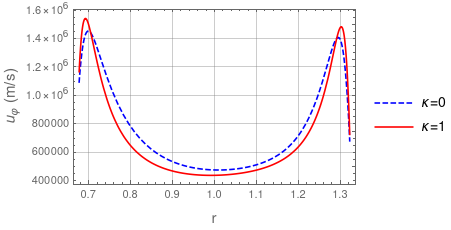}
    \includegraphics[scale=0.45]{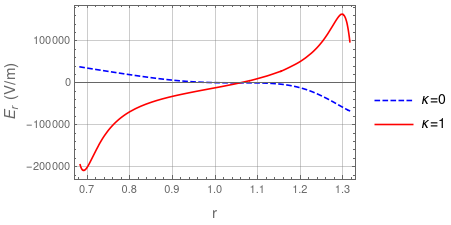}
    \caption{The toroidal rotation velocity profile (left) and the corresponding profile of the $r$ component of the electric field (right) on $z=0$.}
    \label{fig_vt_Er_profiles}
\end{figure}

\begin{figure}[h!]
    \centering
   \hspace{-14mm} \includegraphics[scale=0.45]{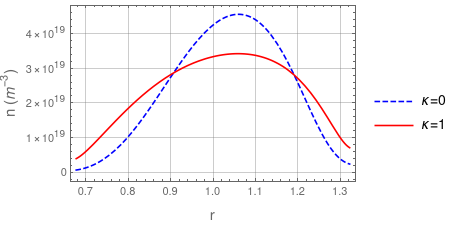} 
    \includegraphics[scale=0.45]{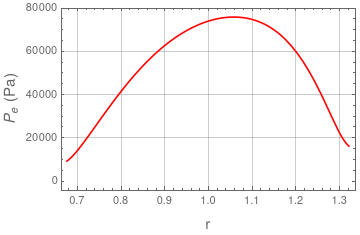}
    \caption{Particle density profiles for $\kappa=0$ and $\kappa=1$ (left panel) and the electron pressure for $\kappa=1$ (right panel).}
    \label{fig_n_profiles}
\end{figure}

\begin{figure}[h!]
    \centering
    \includegraphics[scale=0.45]{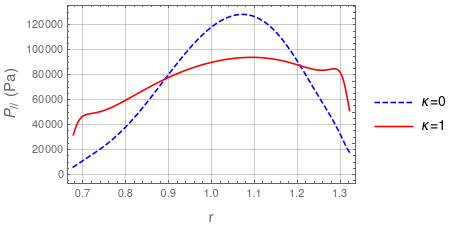}
    \includegraphics[scale=0.45]{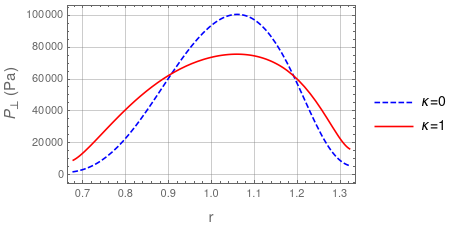}\\
    \caption{Variation of the parallel (left panel) and perpendicular (right panel) components of the ion pressure tensor along $r$ axis on $z=0$ for both $\kappa=0$ and $\kappa=1$.}
    \label{fig_P_profiles}
\end{figure}

\begin{figure}[h!]
    \centering
    \includegraphics[scale=0.35]{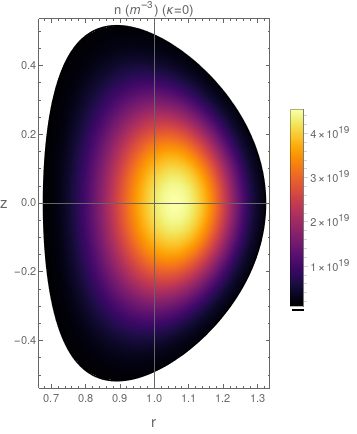}
    \includegraphics[scale=0.35]{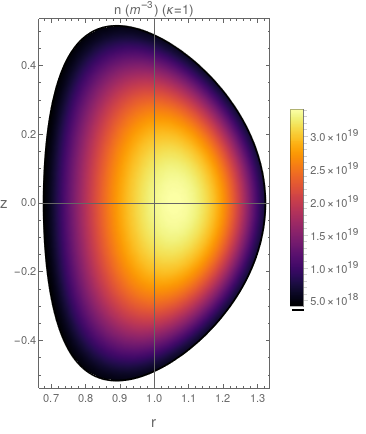}
    \caption{Two dimensional density plots for the particle densities in $\kappa=0$ (left panel) and $\kappa=1$ (right panel) case.}
    \label{fig_dp_n}
\end{figure}

\begin{figure}[h!]
    \centering
    \includegraphics[scale=0.35]{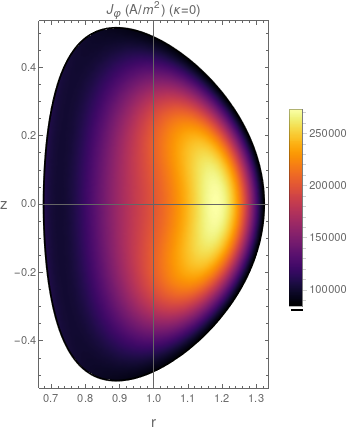}
    \includegraphics[scale=0.35]{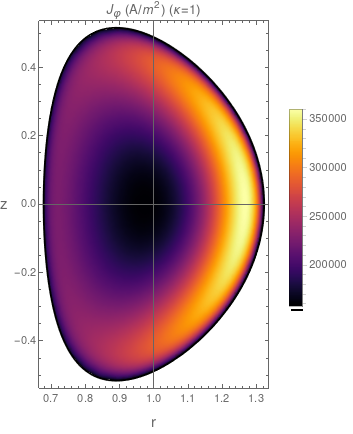}
    \caption{Variation of $J_\phi$ component on the $r-z$ plane for $\kappa=0$ (left) and $\kappa=1$ (right). }
    \label{fig_dp_J}
\end{figure}

\begin{figure}[h!]
    \centering
    \includegraphics[scale=0.35]{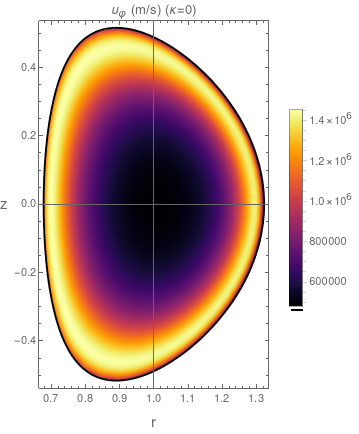}
    \includegraphics[scale=0.35]{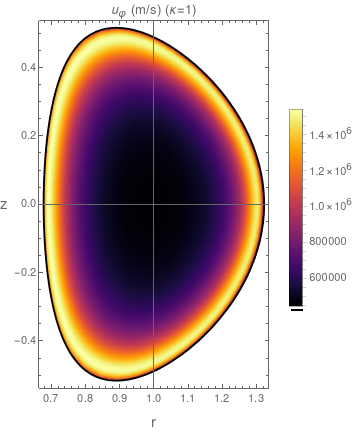}
    \caption{The variation of the toroidal rotation velocity $u_\phi$ on the plane $r-z$, for cold (left) and thermal electrons (right).}
    \label{fig_dp_v}
\end{figure}

\begin{figure}[h!]
    \centering
    \includegraphics[scale=0.35]{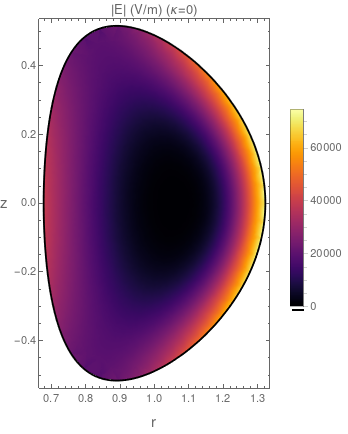}
    \includegraphics[scale=0.35]{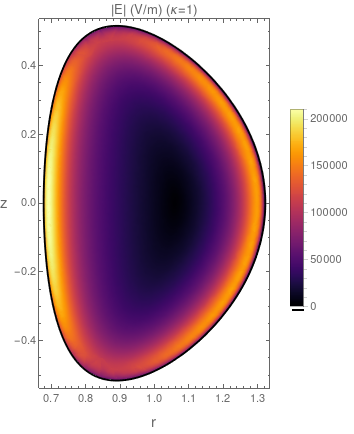}
    \caption{The variation of the electric field magnitude $|\textbf{E}|$ on the plane $r-z$, for cold (left) and thermal electrons (right).}
    \label{fig_dp_E}
\end{figure}

\begin{figure}[h!]
    \centering
    \includegraphics[scale=0.35]{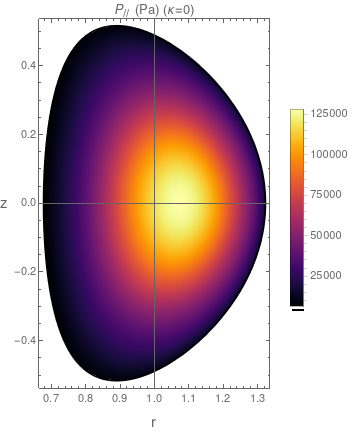}
    \includegraphics[scale=0.35]{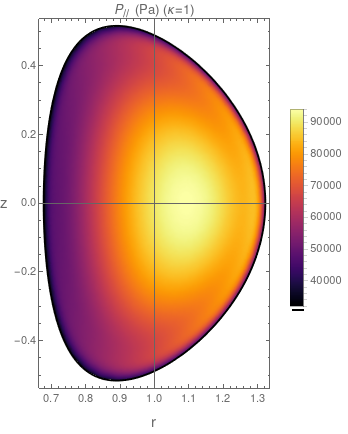}
    \caption{The parallel component ($P_\parallel$) of the ion pressure tensor for cold (left panel) and thermal (right panel) electrons.}
    \label{fig_dp_Ppar}
\end{figure}

\begin{figure}[h!]
    \centering
    \includegraphics[scale=0.35]{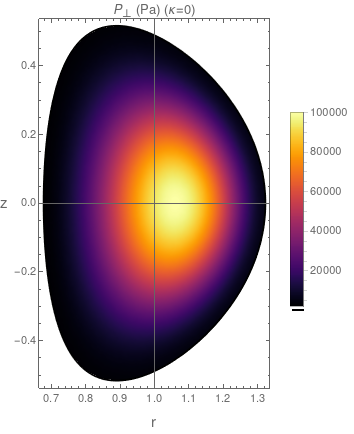}
    \includegraphics[scale=0.35]{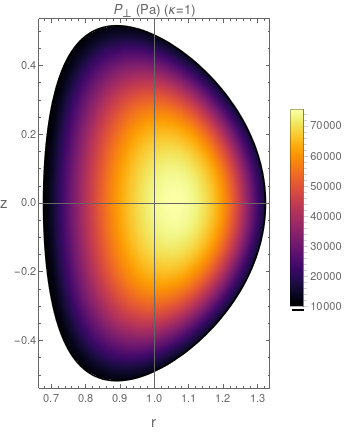}
    \caption{The perpendicular component ($P_\perp$) of the ion pressure tensor for cold (left panel) and thermal (right panel) electrons.}
    \label{fig_dp_Pperp}
\end{figure}

\begin{figure}[h!]
    \centering
    \includegraphics[scale=0.35]{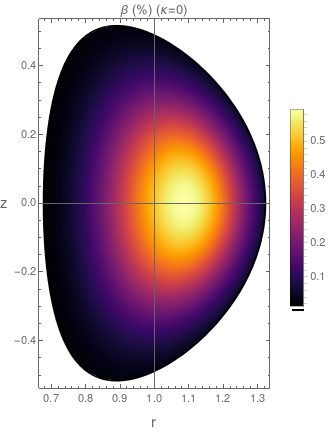}
    \includegraphics[scale=0.35]{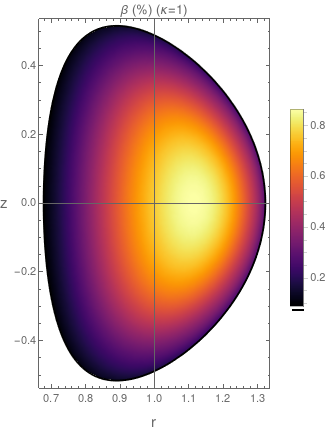}
    \caption{Plasma $\beta$ calculated using \eqref{beta} for the case $\kappa=0$ (left) and the case $\kappa=1$ (right).}
    \label{fig_dp_beta}
\end{figure}

\begin{figure}[h!]
    \centering
    \includegraphics[scale=0.35]{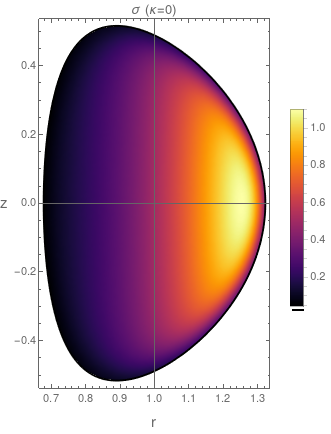}
    \includegraphics[scale=0.35]{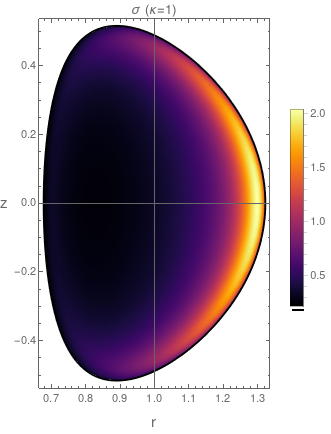}
    \caption{The anisotropy function $\sigma$, calculated by \eqref{sigma} for $\kappa=0$ (left) and $\kappa=1$ (right).}
    \label{fig_dp_sigma}
\end{figure}

\begin{figure}[h!]
    \centering
    \includegraphics[scale=0.37]{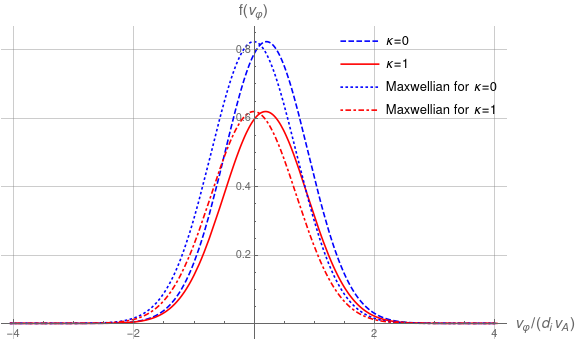}
    \includegraphics[scale=0.37]{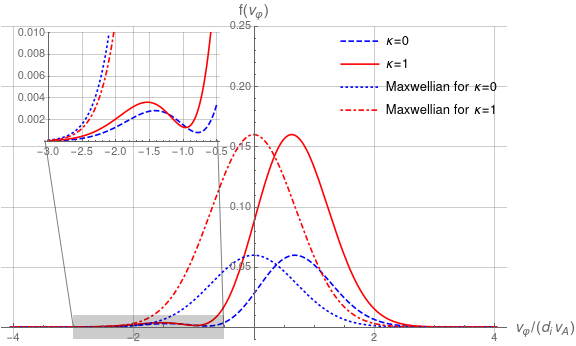}
    \caption{Variation of the ion distribution function $f$ computed by \eqref{DF_special} with $v_\phi$, at two different locations: at the magnetic axis (left) and at an edge point with coordinates ($r=1.3,z=0.0$) (right).}
    \label{fig_df}
\end{figure}

We solve both boundary value problems, corresponding to $\kappa=0$ and $\kappa=1$, using the Finite Element Method (FEM), which is conveniently implemented with Mathematica. The boundary $\partial \mathcal{D}$  is defined as a polygon with a large number of vertices. The vertex coordinates can be boundary points extracted by some parametric formula or by experimental data. The boundary is characterized by an inverse aspect ratio $\epsilon=0.32$, triangularity $\delta=0.34$ and elongation equal to $1.6$. The characteristic values of length, magnetic field and number density used for unit recovery are, respectively $R_0=6.2\,m$, $B_0=5\, T$ and $n_0=2.1\times 10^{19}\, m^{-3}$.  The algorithm performs several iterations to find the position of the magnetic axis which is required for determining the function $G=\alpha(\psi-\psi_a)^2$, until the convergence criterion $max(|\psi_{new}-\psi_{old}|)<\epsilon_{tol}$ is satisfied. For our calculations we have set $\epsilon_{tol}=10^{-7}$. 

The contours of constant $\psi$ (magnetic surfaces) for both equilibria are shown in Fig.~\ref{fig_contours}. These equilibria are calculated by solving \eqref{GS_2} with the ansatz \eqref{ansatz_I} for $I^2(\psi)$. In the case of cold electrons, the function $V$ is given by \eqref{ansatz_V_cold}, while for thermal electrons, $V$ is specified by \eqref{ansatz_V_hot}. The values of the free parameters in the functions $I$ and $V$ are identical for both cases. For the specific examples presented here, we have chosen $I_0=0.5$, $I_1=10^{-1}$, $I_2=-14$, $V_1=15$, $V_2=1.02\times 10^4$, $\alpha=7.5$, and $\eta=0.1$.

The characteristics of the equilibrium can be deduced from Figures \ref{fig_u_Bz} to \ref{fig_P_profiles}, which display variations of various physical quantities of interest along the $r$ axis on the $z=0$ plane. Two-dimensional density plots of the same quantities are presented in Figures \ref{fig_dp_n} to \ref{fig_dp_sigma}. Notably, the particle density in both equilibria does not vanish at the boundary (Figures \ref{fig_n_profiles} and \ref{fig_dp_n}), implying that this equilibrium model is suitable for describing internal plasma regions bounded by a closed magnetic surface which defines the computational domain and does not coincide with the actual plasma boundary.

Additionally, we observe that the toroidal plasma rotation velocity profile exhibits a hollow shape with significant flow shear and radial electric field ($E_r$) in the plasma edge (Figs.~\ref{fig_vt_Er_profiles}, \ref{fig_dp_v}, and \ref{fig_dp_E}). Such edge sheared flows have been associated with the reduction of radial turbulent transport and the transition to high (H) confinement modes in large tokamaks (e.g., \cite{Burrell2020, Plank2023}). Moreover, the toroidal current density profile for the $\kappa=1$ equilibrium shows a reduction in the central region of the plasma (Figs.~\ref{fig_J_profiles}, \ref{fig_dp_J}). 

We also examine the parallel and perpendicular components of the ion pressure tensor, as presented in Figures \ref{fig_dp_Ppar} and \ref{fig_dp_Pperp}. The steps for obtaining $P_{\parallel}$ and $P_\perp$ can be found in Appendix \ref{appendix_B}. In Fig.~\ref{fig_P_profiles}, we provide profiles of these components on the plane $z=0$. Notably, the $P_\parallel$ component of the ion pressure tensor forms a pedestal in the thermal electron case. As a consequence, the effective pressure defined as  $(P_\parallel+P_\perp)/2$ also forms a pedestal due to the $P_\parallel$ contribution.

In addition to the previously mentioned physical quantities, we calculate two figures of merit for both cold and thermal electron equilibria: the plasma $\beta$ and the anisotropy function $\sigma$ (defined in Appendix \ref{appendix_B}, Eq.~\eqref{sigma}). In nondimensional form the expression for calculating the plasma $\beta$ is:
\begin{eqnarray}
    \beta = d_i^2\frac{P_e + \langle P \rangle} {B^2}\,, \label{beta}
\end{eqnarray}
where $\langle P \rangle := (P_{rr}+P_{zz}+P_{\phi\phi})/3$ with $P_{rr},P_{zz}, P_{\phi\phi}$ being the diagonal components of the pressure tensor (see Appendix \ref{appendix_B}). The presence of the $d_i^2$ factor arises owing to the specific scaling we have adopted for the normalized pressure in Eqs.~\eqref{normalization}. Figures \ref{fig_dp_beta} and \ref{fig_dp_sigma} illustrate that the plasma $\beta$ ranges from approximately $0.5-1.0\%$ and increases from the plasma boundary towards the core, while the ion pressure anisotropy is more pronounced on the low-field side of the configuration.

We conclude our presentation of equilibrium results with Fig.~\ref{fig_df}, which illustrates the variation of ion distribution functions as a function of the toroidal particle velocity component $v_\phi$, for both $\kappa=0$ and $\kappa=1$ cases at two distinct locations: the magnetic axis $(r_{ax},z_{ax})$ and an edge point with coordinates $(r=1.3,z=0.0)$. In both cases, the dependence on $v_r$ and $v_z$ has been eliminated by integrating the distribution functions over the $v_r-v_z$ plane. The two distribution functions are presented alongside the corresponding normalized Maxwellian distributions $f_0 e^{-v_\phi^2}$, where $f_0$ is an appropriate normalization constant. In both cases, the distributions exhibit a shift towards positive $v_\phi$, resulting in finite macroscopic toroidal flows. At the edge point $(r=1.3,z=0)$, where the toroidal flow appears to reach a maximum, the distributions significantly deviate from the Maxwellian, displaying a bump-on-tail form. The bump corresponds to ions rotating in the opposite direction of the macroscopic flow.

\section{Summary}
\label{Sec_V}
In this work, we have presented the axisymmetric equilibrium formulation of the hybrid Vlasov equilibrium model introduced in \cite{Kaltsas2023}, featuring massless electrons and kinetic ions. We derived a general form of the Grad-Shafranov equation and outlined a method for determining ion distribution functions in terms of Hermite polynomials based on the knowledge of the total and the electron current density profile. Our formulation allowed us to solve the equilibrium problem for specific choices of the arbitrary functions involved in the Grad-Shafranov equation. The results demonstrate the model's capability to describe plasmas with geometric and profile characteristics relevant to tokamaks. Notably, these equilibria exhibit some features reminiscent of H-mode phenomenology, including strongly sheared edge flows and significant edge radial electric fields. Building upon these results, more refined descriptions of plasma equilibria with kinetic effects stemming from kinetic particle populations are possible. Thus, future research will focus on improving the model to incorporate realistic electron temperature distribution and fluid ion components. An intriguing open question is whether this equilibrium model can be derived through a Hamiltonian energy-Casimir (EC) variational principle, as explored in \cite{Kaltsas2021,Tronci2015}. Identifying the complete set of Casimir invariants of the dynamical system is crucial for such a variational formulation of the equilibrium problem and for establishing stability criteria within the Hamiltonian framework. Note that, in general, there are not enough Casimirs to recover all the possible classes of equilibria due to rank changing of the Poisson operator (see \cite{Morrison1998}). However, instead of the EC variational principle, one can apply an alternative Hamiltonian variational method that recovers all equilibria upon utilizing dynamically accessible variations \cite{Morrison1998}.

\section*{Acknowledgements}
This work has received funding from the National Fusion Programme of the Hellenic Republic – General Secretariat for Research and Innovation. P.J.M. was supported by the U.S. Department of Energy Contract No. DE-FG05-80ET-53088.

\begin{appendices}

\section{Expansion of $g(p_\phi)$ in terms of Hermite polynomials}
\label{appendix_A}
When the product $V^{\kappa+1} e^{G(\psi)/d_i^2}$ can be expressed as a power series expansion of $\psi$, it is possible to determine the function $g(p_\phi)$ in terms of Hermite polynomials. To see this, we invoke that Hermite polynomials $H_n(x)$ serve as coefficients in the following power series expansion \cite{Morse1953}:
\begin{eqnarray}
    e^{-(x-y)^2/2} = \sum_{n=0}^{\infty} \frac{e^{-x^2/2}}{n!} H_n\left(\frac{x}{\sqrt{2}}\right)\left(\frac{y}{\sqrt{2}}\right)^n\,,
\end{eqnarray}
 therefore \eqref{integral_V} can be written as
\begin{eqnarray}
    \left[\frac{V}{d_i^2(\kappa+1)}\right]^{\kappa+1} e^{G(\psi)/d_i^2} = \sum_{n} \frac{2\pi}{n!} \int_{-\infty}^{+\infty} d\zeta\, e^{-\zeta^2/2}H_n\left(\frac{\zeta}{\sqrt{2}}\right)\left(\frac{\psi}{d_i^2\sqrt{2}r}\right)^ng(r \zeta)\,, \label{V=sum-Hermite_1}
\end{eqnarray}
where $\zeta:= p_\phi/r$. As Hermite polynomials form a complete orthogonal basis, we can expand $g(r\zeta)$ as
\begin{eqnarray}
    g(r\zeta) = \sum_m c_m H_m\left(\frac{r\zeta}{\sqrt{2}}\right)\,. \label{g_Hermite}
\end{eqnarray}
We now make use of the multiplication theorem for Hermite polynomials \cite{Chaggara2007}
\begin{eqnarray}
    H_m(\gamma x) = \sum_{\ell=0}^{\lfloor m/2 \rfloor} \gamma^{m-2\ell}(\gamma^2-1)^\ell \frac{m!}{\ell !(m-2\ell)!}H_{m-2\ell}(x) \,, \quad \forall \gamma \in \mathbb{R}\,,
\end{eqnarray}
to write
\begin{eqnarray}
    g(r\zeta) = \sum_{m} \sum_{\ell=0}^{\lfloor m/2 \rfloor} c_m \frac{m!}{\ell !(m-2\ell)!} r^{m-2\ell}(r^2-1)^\ell H_{m-2\ell}\left(\frac{\zeta}{\sqrt{2}}\right)\,. \label{g_Hermite_dupl_theorem}
\end{eqnarray}
Substituting \eqref{g_Hermite_dupl_theorem}, the right hand side (rhs) of \eqref{V=sum-Hermite_1} becomes
%\begin{eqnarray}
%    \left[\frac{V}{d_i^2(\kappa+1)}\right]^{\kappa+1}e^{G(\psi)/d_i^2} = \sum_{m,n} \frac{2\pi}{n !} c_m \int_{-\infty}^{+\infty}d\zeta\, e^{-\zeta^2/2}H_n\left(\frac{\zeta}{\sqrt{2}}\right)H_m\left(\frac{r\zeta}{\sqrt{2}}\right)\left(\frac{\psi}{d_i^2\sqrt{2}r}\right)^n\,.
%\end{eqnarray}
%
\begin{eqnarray}
\sum_{m,n}\sum_{\ell=0}^{\lfloor m/2 \rfloor}\frac{2\pi}{n !} c_m \frac{m!}{\ell !(m-2\ell)!} r^{m-2\ell}(r^2-1)^\ell \left(\frac{\psi}{d_i^2\sqrt{2}r}\right)^n \times \nonumber \\
    \times  \int_{-\infty}^{+\infty}d\zeta\, e^{-\zeta^2/2}H_n\left(\frac{\zeta}{\sqrt{2}}\right)H_{m-2\ell}\left(\frac{\zeta}{\sqrt{2}}\right)\,. \label{V=sum-Hermite_2}
\end{eqnarray}
Further, exploiting the orthogonality condition
\begin{eqnarray}
    \int_{-\infty}^{+\infty} dx \, H_n(x)H_m(x) e^{-x^2}= \sqrt{\pi}2^n n! \delta_{mn}\,,
\end{eqnarray}
we can see that Eq.~\eqref{V=sum-Hermite_1} with rhs given by \eqref{V=sum-Hermite_2}, becomes
\begin{eqnarray}
        \left[\frac{V}{d_i^2(\kappa+1)}\right]^{\kappa+1} e^{G(\psi)/d_i^2}= \sum_{m}\sum_{\ell=0}^{\lfloor m/2 \rfloor}2^{m-2\ell+1}\pi^{3/2} c_m \frac{m!}{\ell !(m-2\ell)!} (r^2-1)^\ell \left(\frac{\psi}{d_i^2\sqrt{2}}\right)^{m-2\ell}\,. \label{V=sum-Hermite_3}
\end{eqnarray}

\section{Calculation of the ion pressure tensor components}
\label{appendix_B}
The ion pressure tensor is defined by
\begin{eqnarray}
    \textbf{P} = \int d^3v\, (\bsv - \textbf{u})(\bsv-\textbf{u}) f\,,
\end{eqnarray}
where $\textbf{u}$ is calculated by \eqref{u_integral}. In our case $\textbf{u} = u_\phi \hat{\phi}$. Selecting the $(v_r, v_z, v_\phi)$ basis in the velocity space we calculate below the following diagonal pressure components $P_{rr},P_{zz}, P_{\phi}$. Note that the non-diagonal components $P_{rz}=P_{r\phi}=P_{z\phi}=0$ vanish owing to the fact that $f$ is an even function of the velocity components $v_r$, $v_z$. The diagonal elements are calculated as follows
\begin{eqnarray}
    P_{rr} &=& \int d^3v \, v_r^2 f \,,\\
    P_{zz} &=& \int d^3v \, v_z^2 f\,, \\
    P_{\phi\phi} &=& \int d^3v\, (v_\phi-u_\phi)^2 f\,.
\end{eqnarray}
An average value of the ion pressure is given by $\langle P \rangle = Tr(P_{ij})/3$.

It is evident that $P_{rr}=P_{zz}$ and since the non-diagonal components are zero, the ion pressure tensor is gyrotropic and can be written in the form 
\begin{eqnarray}
    \textbf{P} = \sigma \textbf{B}\textbf{B} +P_\perp \textbf{I}\,, \label{sigma}
\end{eqnarray}
where 
$$\sigma \coloneqq d_i^2\frac{P_{\parallel} - P_\perp}{B^2}\,,$$
is an anisotropy function. Note that the factor $d_i^2$ appears due to the specific scaling of the particle velocity adopted in \eqref{normalization}. The parallel and the perpendicular to $\textbf{B}$ components of the pressure tensor can be calculated by the following relations
\begin{eqnarray}
    P_{\parallel} &=& \frac{\textbf{P} \boldsymbol{:} \textbf{B}\textbf{B}}{B^2} = \frac{P_{ij}B_{i}B_j}{B^2}\,, \\
    P_\perp &=& \frac{1}{2}\textbf{P}\boldsymbol{:}\left(\textbf{I}-\frac{\textbf{B}\textbf{B}}{B^2}\right)=\frac{1}{2}P_{ij}\left(\delta_{ij}- \frac{B_iB_j}{B^2}\right)\,.
\end{eqnarray}

\end{appendices}
%\section*{References} 
\printbibliography

\end{document}